\documentclass[5p,twocolumn,10pt]{elsarticle}
\usepackage{graphics}
\usepackage{amsmath,amsthm,amsfonts,amssymb}
\usepackage{amssymb}

\journal{Physica C}
\begin{document}
\begin{frontmatter}

\title{Study of higher-order harmonics of complex ac susceptibility in $\mathrm{Y}\mathrm{B}_2\mathrm{C}_3\mathrm{O}_{7-\delta}$ thin films by the mutual inductive method}

\author[label1]{Israel P\'erez\corref{cor1}}
\cortext[cor1]{Corresponding author}
\ead{iperez@mda.cinvestav.mx}
\author[label1]{V\'ictor Sosa}
\address[label1]{Applied Physics Department,
Cinvestav Unidad M\'erida, Km 6 Ant., Carretera a Progreso, A.P. 73, C.P. 97310. M\'erida, Yucat\'an M\'exico}
\author[label2]{Fidel Gamboa}
\ead{ffgamboa@mda.cinvestav.mx}
\address[label2]{Physics of Materials Department, Centro de Investigaci\'on en Materiales Avanzados, S.C. (CIMAV) Ave. Miguel de Cervantes 120, Complejo Industrial Chihuahua, Chihuahua, M\'exico}

\begin{abstract}
We have applied the mutual inductive method to study higher-order harmonics of complex ac susceptibility $\chi_n=\chi'_n-i\chi''_n$ for $\mathrm{Y}\mathrm{B}_2\mathrm{C}_3\mathrm{O}_{7-\delta}$ thin films as function of the temperature and the applied field. The experimental results were compared with analytical and numerical results obtained from the Ishida-Mazaki model and the solution of the integral equation for the current density, respectively. Both models allow us to reproduced the main experimental features, however, as $n$ increases the numerical model shows notable discrepancies. This failure can be attributed to the current-voltage characteristics. Also this investigation yields the activation energy $U_c$ and the critical current density $J_c$ for two samples both at $T=0$.
\end{abstract}

\begin{keyword}
Mutual Inductive Method; Harmonics; Thin films
\PACS{74.25.Ha,74.25.Qt,74.25.Sv,74.25.Nf, 74.78.Bz}
\end{keyword}

\end{frontmatter}

\section{Introduction}

Since the pioneer works of Bean \cite{bean1} the analysis of the harmonics of the complex ac susceptibility $\chi_n=\chi'_n-i\chi''_n$ has become one of the most common methods used for the characterization of high-temperature superconductors \cite{chen2,woch}. Conventionally, the ac susceptibility technique \cite{nikolo3} is used in most physics laboratories; however, the inductive method has shown to be an alternative and highly efficient method not only to study the electromagnetic properties and the vortex dynamics in superconducting thin films but also some important aspects of the models \cite{classen,yamada,acosta3,acosta6,perez3}. The physical information contained in the harmonics of susceptibility have been extensively studied \cite{acosta3,acosta6,gomory,valenzuela}. The real part of the fundamental susceptibility $\chi'$ is directly related to the shielding supercurrents (or diamagnetic response) flowing in the film; it gives information on how the magnetization in the sample is evolving; whereas the energy dissipation occurring in the sample is displayed by the imaginary component $\chi''$. Moreover, it is well known that the irreversibility of magnetization is the signature for trap flux within the sample. The nonlinear magnetic response of the superconductor when flux pinning is present, is reflected in the presence of higher harmonics. Great efforts have been made to understand the physics implied in these harmonics not only for superconductors in bulk \cite{ishida2, ishida,qin2,ozogul,moreno,adesso1,adesso2} but also for superconducting thin films \cite{perez3,aruna}. The third harmonic $\chi_3$ has shown to be an indicative of flux penetration into the sample and has been embraced as a criterion for the determination of the critical current density $J_c$ \cite{classen,yamada,acosta3,perez3}. In a similar fashion, the decrease in the value of higher harmonics has been related to a reduction of the oxygen contents in $\mathrm{Y}\mathrm{B}_2\mathrm{C}_3\mathrm{O}_{7-\delta}$ crystals \cite{moreno}. Other works \cite{perez3} have suggested that models based solely on the critical state model \cite{bean1,sun1} cannot reproduce the experimental data of the higher harmonics due to thermal effects like flux creep. 

One of the motivations of this paper is to test the validity of some of the present models on the one hand, and determine the values of their parameters on the other. For this purpose we aport a study of the higher harmonics for the temperature and magnetic field dependence of the complex ac susceptibility ($\chi_n=\chi'_n-i\chi''_n,\; n=3,5$) in $\mathrm{Y}\mathrm{B}_2\mathrm{C}_3\mathrm{O}_{7-\delta}$ thin films. The ac field is applied normal to the sample surface between $0$ and $3500$ A/m and the temperature is varied, for each run, from 70 to 82 K. Then, the collected experimental data is compared with the Brandt model \cite{brandt11,brandt13,brandt14} which deals with the integral equation for the current density in two dimensions. Furthermore, a comparative analysis of the well established Ishida-Mazaki model \cite{ishida}, which is a good approximation for inhomogeneous superconductors, is carried out. This investigation also allows us to estimate the activation energy $U_c$ and the critical current density $J_{c}$ both at $T=0$. The results of the study reveals that the Ishida-Mazaki model reproduces better the experimental curves. This suggests that the microstructure of thin films behaves as a set of layers forming multiconnected network of superconducting Josephson weak links.

\section{Theoretical Models}
In order to interpret our data, we compared them with numerical and analytical models. In this article we have worked out the numerical calculations derived from the solution of the non-linear integral equation for the current density which allows us to analyze the dynamics of the critical current density whenever there is a nontrivial dependence on the position. Also we have adopted the analytical expression for the magnetic susceptibility of Ishida-Mazaki model which was developed for the study of inhomogeneous superconductors.

In geometrical terms we considered a thin disc of radius $a$ and thickness $d=2b$ immersed in an applied time-dependent magnetic field $B_a(t)=B_{ac}\cos (\omega t)$ perpendicular to the surface of the film, where $B_{ac}$ is the amplitude of the alternating field. These assumptions are valid for both models; the main differences are now readily established.

\subsection{Numerical model: current density equation}
 Following Brandt  \cite{brandt11, brandt13,brandt14}, besides the above considerations, we also assumed a nonlinear response of the current-voltage characteristics, i.e., $\mathbf{E}=\rho(J)\mathbf{J}$. The problem, then, reduces to find the solution of the integral equation for our geometry:
\begin{equation}
\label{otj}
 J(r,t)=-\mu_0^{-1}\int_0^ad^2r'Q^{-1}(r,r')[A(r',t)+\frac{r}{2}B_a(t)].
\end{equation}
where $\dot{A}(r,t)  = -E[J(r',t)]$ and $\mu_0$ is the magnetic susceptibility for the vacuum and the total vector potential is given by 
\begin{equation}
\label{pot1}
A(\mathbf{r},t)=-\mu_0\int_0^adr'\int_0^b dy'Q(\mathbf{r},\mathbf{r}')J(\mathbf{r}',t) -\frac{r}{2}B_a(t)
\end{equation}
where $\mathbf{r}(r,y)$, $\mathbf{r}'(r,y)$. The integrals run over the spatial dimensions of the sample and
\begin{equation}
\label{g2}
Q(r,r')=\int_0^{\pi}\frac{-r'\cos \varphi'\, d\varphi'}{[r^2+r'^2-2rr'\cos \varphi']^{1/2}}.
\end{equation}
is the integral kernel.

\subsection{Material law}

From the theories of collective creep, flux creep and also from the vortex glass picture \cite{kim1,kim2,acosta4,feigelman,nattermann,fisher}, the material law equation $E(J)$ becomes strongly nonlinear and can be generally described by 
\begin{equation}
\label{matlaw}
E(J)=E_c\exp\biggl[\frac{-U(J)}{k_BT}\biggr],
\end{equation}
where $k_B$ is the Boltzman constant, $T$ is the temperature of the sample, the prefactor $E_c=B_av_0$ and $v_0$ is the flux creep velocity. Next we introduce the current density $J$-dependent activation energy
\begin{equation}
\label{udj}
U(J)=\frac{U_c}{\mu}\biggl[\biggl(\frac{J_c}{J}\biggr)^{\mu}-1\biggr]
\end{equation}
for the depinning, where $U_c$ is a characteristic energy scale and $J_c(T,B)$ is the critical current density which could be function of the magnetic field and the temperature. The glassy exponent $\mu$ is a universal constant in the vortex-glass theory with a value $\sim1$. If $\mu \to 0$ then $U(J)\approx U_c\ln (J_c/J)$. Introducing this into equation \eqref{matlaw} we obtain the power law \cite{brandt13}:
\begin{equation}
\label{ej2}
E(J)=E_c\biggl(\frac{J}{J_c}\biggr)^{m},
\end{equation}
where $m=U_c/(k_BT)$ is a parameter known as the creep exponent valid for the interval $1\le m<\infty$. Typically creep behavior is for the interval $1< m < 20$ \cite{valenzuela}. The nonlinear resistivity for $m\gg 1$ is caused by thermally activated depinning with the activation energy \eqref{udj}. The case $m=1$ is the resistive or flux flow regime, which leads to normal resistivity (Ohmic behavior) and, $m\to \infty$ reduces to the Bean model. 

The temperature dependence of  $J_c$ may be modeled by the well known empirical relation \cite{acosta6}:
\begin{equation}
\label{jemp}
J_c(\tau)=J_{c0}(1-\tau)^r,
\end{equation}
where $\tau=T/T_c$ is the reduced temperature, $T_c$ is the critical temperature, $J_{c0}$ is the current density at $\tau=0$ and $r$ is a fitting parameter. We also tried other temperature dependences for the critical current density, for instance, within the context of the Ginzburg-Landau theory the critical current is given by $J_c(\tau)=J_{c0}(1-\tau^2)^{5/2}/(1+\tau^2)^{1/2}$, and for the two fluid model we have $J_c(\tau)=J_{c0}(1-\tau^2)(1-\tau^4)^{1/2}$. However, we find that the higher harmonic response is well described with equation \eqref{jemp}.

On the other hand, the characteristic energy within the collective creep model can be described by
\begin{equation}
\label{uctb}
U_c(\tau)=U_{c0}(1-\tau^4),
\end{equation}
where $U_{c0}$ is the activation energy at $\tau=0$. With the preceding information at hand we could estimate the creep parameter. For the temperature range treated in this work the creep parameter varies from 1.2 for 82 K to 6 for 70 K.

One of the advantages of the power law \eqref{ej2} is that the problem reduces to the manipulation of only two parameters, that is, $U_{c0}$ and $J_{c0}$. The initial values for $U_{c0}$ and $J_{c0}$ were taken from the literature \cite{perez3,moreno,aruna} (see next section), however, once that the first computation is given, we fixed one parameter, say $J_{c0}$, and manipulate the other. By comparison with the experimental results we obtained the best values of $U_{c0}$ for each sample. 

\subsection{Numerical Integration}
\label{numint}

The time integration of equation \eqref{otj} has to be performed numerically \cite{brandt14}. This may be done by tabulating the functions $J$, $E$, and $A$ on a 2-D non-equidistant grid. A possible nonequidistant grid $\mathbf{r}_i=(r_i,y_i)$ is obtained by the sustitutions $r=r(u)=1/2(3u-u^3)a$, $y=y(v)=1/2(3v-v^3)b$, and then tabulating $u=0,\dots, 1$ and $v=0,\dots, 1$ on equidistant grids $u_k=(k-1/2)/N_r$ ($k=1,\dots,N_r$) and $v_l=(l-1/2)/N_y$ ($l=1,\dots,N_y$); this yields a 2-D grid of $N=N_rN_y$ points with weights $w_i=w_rw_y$, $w_r=dr_k/dk=3/2(1-u_k^2)a/N_r$, and $w_y=dy_l/dl=3/2(1-v_l^2)b/N_y$, which vanishes at the boundaries $r=a$ and $y=b$. Labeling the points $(r_k,y_l)$ by one index $i=1,2,\dots,N$ the functions $J(r,y,t)$ become time dependent vectors $J_i(t)$ with $N$ components and the integral kernel becomes a $N\times N$ matrix $Q_{ij}$. Thus, equation \eqref{otj} yields the following discrete expression
\begin{equation}
\label{je2}
J_i(t)=-\mu^{-1}_0\frac{b}{N}\sum_{j=1}^N(Q_{ij}w_i)^{-1}\psi_j,
\end{equation}
with
\begin{equation}
\label{psidis}
\psi_j=A_j(t)+\frac{r_j}{2}B_a(t).
\end{equation}
This equation can be solved by a Four-order Runge-Kutta method or by a step-by-step numerical integration; both numerical methods give stable results. In our case the space variables were discretized with $N=450$ non-equidistant grid points. We used the initial values $U_{c0}=480$ K and $J_{c0}=3\times 10^{11}\, \textrm{A}/\textrm{m}^2$. However, after the first computation, we compared the results with the experimental ones. We found that the best fitting values for $U_{c0}$ were 930 K and 1300 K for the samples M22 and M23, respectively (see next sections). It is to be seen that these values were kept fixed during the subsequent computations, i.e., for the different temperatures.

Having obtained the sheet current density the total magnetic moment per unit length of a disk was calculated using the expression
\begin{equation}
\label{magdisk}
M=\int_Sd^2r \mathbf{r}\times \mathbf{J}=2\pi \int_0^ar^2dr\int_0^bdy J(r,y).
\end{equation}
And, finally, the components of the magnetic susceptibility were reckoned using the typical relations:
\begin{eqnarray}
\label{sunsus}
\chi_n' & = & \frac{1}{B_{ac}\pi}\int_0^{2\pi}M(t)\sin(n\omega t) d(\omega t), \nonumber \\
\chi_n'' & = & \frac{1}{B_{ac}\pi}\int_0^{2\pi}M(t)\cos(n\omega t) d(\omega t) \, .
\end{eqnarray}

\subsection{The Ishida-Mazaki Model}
In the Ishida-Mazaki model \cite{ishida1} inhomogeneous superconductors are modeled as a stack of layers forming networks of multi-connected loops closed by Josephson weak links. In such case the susceptibility is analytically given by
\begin{eqnarray}
\label{susishi1}
\chi'_1 & = & -(1/4\pi^2)\Bigl(\alpha-\frac{1}{2}\sin 2\alpha\Bigr), \nonumber \\
\chi''_ 1& = & (1/4\pi^2)\sin^2 \alpha, 
\end{eqnarray}
\begin{eqnarray}
\label{susishin}
\chi'_n  &=& \frac{1}{4n\pi^2}\biggl(\frac{\sin(n+1)\alpha}{n+1}-\frac{\sin(n-1)\alpha}{n-1}\biggr), \nonumber \\
\chi''_n  &=& -\frac{1}{4n\pi^2}\biggl(\frac{\cos(n+1)\alpha-1}{n+1} \\ \nonumber
&&-\frac{\cos(n-1)\alpha-1}{n-1}\biggr), 
\end{eqnarray}
where 
\begin{equation}
\label{eqishi}
\alpha=2\sin^{-1}(\sin \theta)^{1/2}, \qquad \sin \theta=B_{m}/B_{a}. 
\end{equation}
In this case $B_{m}$ is the magnetic field which induces super-current corresponding to $J_c$ in the loop. This field plays the role of the full penetration field $H^*$ within the context of the Bean model \cite{bean1}. Also for $n=$even, $\chi'_n$ and  $\chi''_n$ do not appear. 

On the other hand, we can introduce the temperature dependence of the magnetic susceptibility via $B_m$. If the critical state is present, thus a linear relation between $B_m$ and $J_c$, from equation \eqref{jemp}, can be assumed \cite{perez3}. The parameter $r$ is close to 1 whenever $T$ is more than 10 K below $T_c$ and $\sim 3.5$ for temperatures very close to $T_c$ \cite{perez3}. For the temperature interval of this work the best fitting was achieved when $r=1.5$, whilst $J_{c0}$ was estimated from the experimental curves with $r$ fixed (see section \ref{dis}). 

\section{Experimental}
\label{exper}
Two superconducting thin films of $\mathrm{Y}\mathrm{B}_2\mathrm{C}_3\mathrm{O}_{7-\delta}$ were deposited on $10\,mm \times 10mm$ heated $\mathrm{La}\mathrm{Al}\mathrm{O}_3$ substrates, using the RF sputtering technique. These films were labeled as M22 and M23. Their properties are summarized in Table \ref{proper}. 

 \begin{table}[htp]
   \caption{Properties for the samples used in this work.}
  \centering 
  \begin{tabular}{c|cccc}
  \hline
\footnotesize{Sample} &\footnotesize{$T_c$} &  \footnotesize{$\Delta T$}&  \footnotesize{Thickness} &  \footnotesize{$J_c(70\,K)$}
\\  \hline \hline 
 &  (K)   & (K)  & (nm)  & $\times 10^{10}$ $\textrm{A}/\textrm{m}^2$
  \\ \hline
\footnotesize{M22} & \footnotesize{$86.0 \pm 0.5$} & \footnotesize{$6.0 \pm 0.5$} & \footnotesize{$266 \pm 5$} &  \footnotesize{$2.7\pm 0.5$} \\ 
\footnotesize{M23} & \footnotesize{$86.5 \pm 0.5$} & \footnotesize{$5.0 \pm 0.5$} & \footnotesize{$260 \pm 30$} & \footnotesize{$2.5\pm 0.5$} \\ 
\end{tabular}
\label{proper}
\end{table}

For the detection of harmonic generation we have used the so-called mutual inductive method  \cite{classen,acosta3,acosta6,perez3}. In this technique a multiturn excitation coil is pressed against the film and a pickup coil is coaxially placed on the rear of the film to sense the deformed wave. The diameters of the coils were approximately 5 mm and the film-coil distance was $<1$ mm.  The driving coil generates a 1 KHz transverse ac magnetic field whose amplitude could vary from 0 to 3500 A/m. A SR830 DSP Stanford lock-in amplifier is used to Fourier analyze the deformed signal whose harmonic values are fed to a computer program where data are stored and displayed in real time. To control the sample temperature, films were fixed with silver glue on a Cryomech ST15 cold finger. The temperature measurements were performed with a 9600-1 Scientific Instruments controller and a Si-410NN silicon diode with a resolution of 0.1 K.

\section{Results and discussion}
\label{dis}

We have measured the real and imaginary components of susceptibility response as function of the applied field $B_{a}$ up to the fifth harmonic for the two superconducting films at five different temperatures, namely: 70 K, 75 K, 78 K, 80 K and 82 K. As mentioned in the introduction of this contribution we shall focus on the analysis of harmonics for $n=3,5$. 
 \begin{figure}[htp]
\begin{center}
\includegraphics[width=7cm]{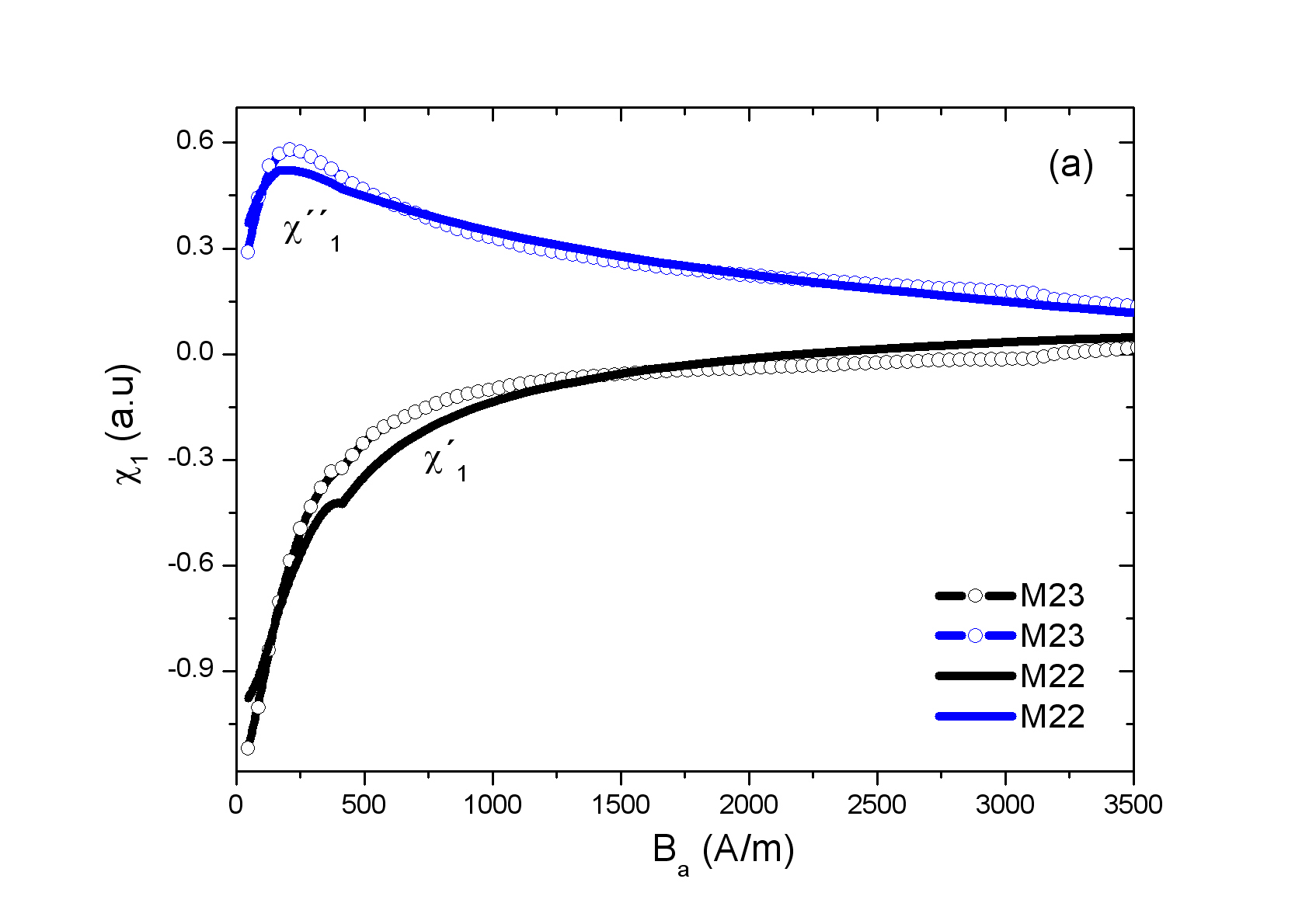} \includegraphics[width=7cm]{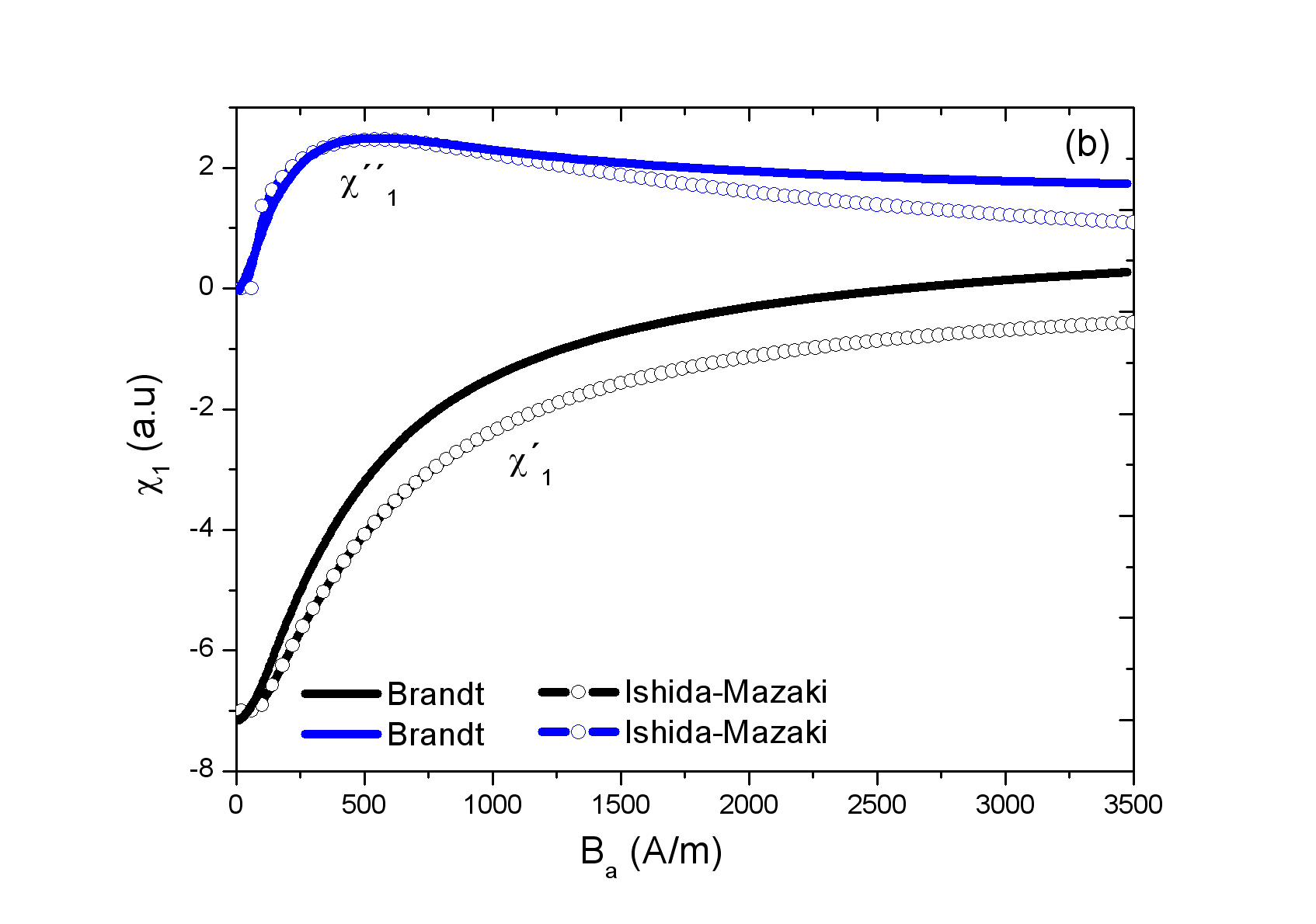} 
\caption{Representative curves for the fundamental harmonic of ac susceptibility response as function of the $B_a$ at 78 K. (a) experimental curves for the two samples used in this work. (b) Ishida-Mazaki calculations and Brandt model. There is an overall agreement between theory and experimental data.}
\label{first}
\end{center}
\end{figure}
 However, in order to verify the consistency of the experimental data among the films and our models we have plotted in Fig. \ref{first} representative curves for the fundamental component of susceptibility at 78 K. Figure \ref{first}a compares the curves for the real and imaginary components for the two samples. We can see that there is a minor difference in the structure of the curves. Figure \ref{first}b compares the curves for the real and imaginary components for the two models used in this work. As the field increases the amplitude of the numerical curves for $\chi''$ falls off slower to zero than that of the Ishida-Mazaki calculations. Similarly, the opposite occurs for $\chi'$. These slight discrepancies arise from the different assumptions for each model. In particular the numerical calculations assume creep effects. The complete analysis for the fundamental harmonic has been extensively study by Acosta et al. \cite{acosta3,acosta6}.

\begin{figure*}[htp]
\begin{center}
\includegraphics[width=6.5cm]{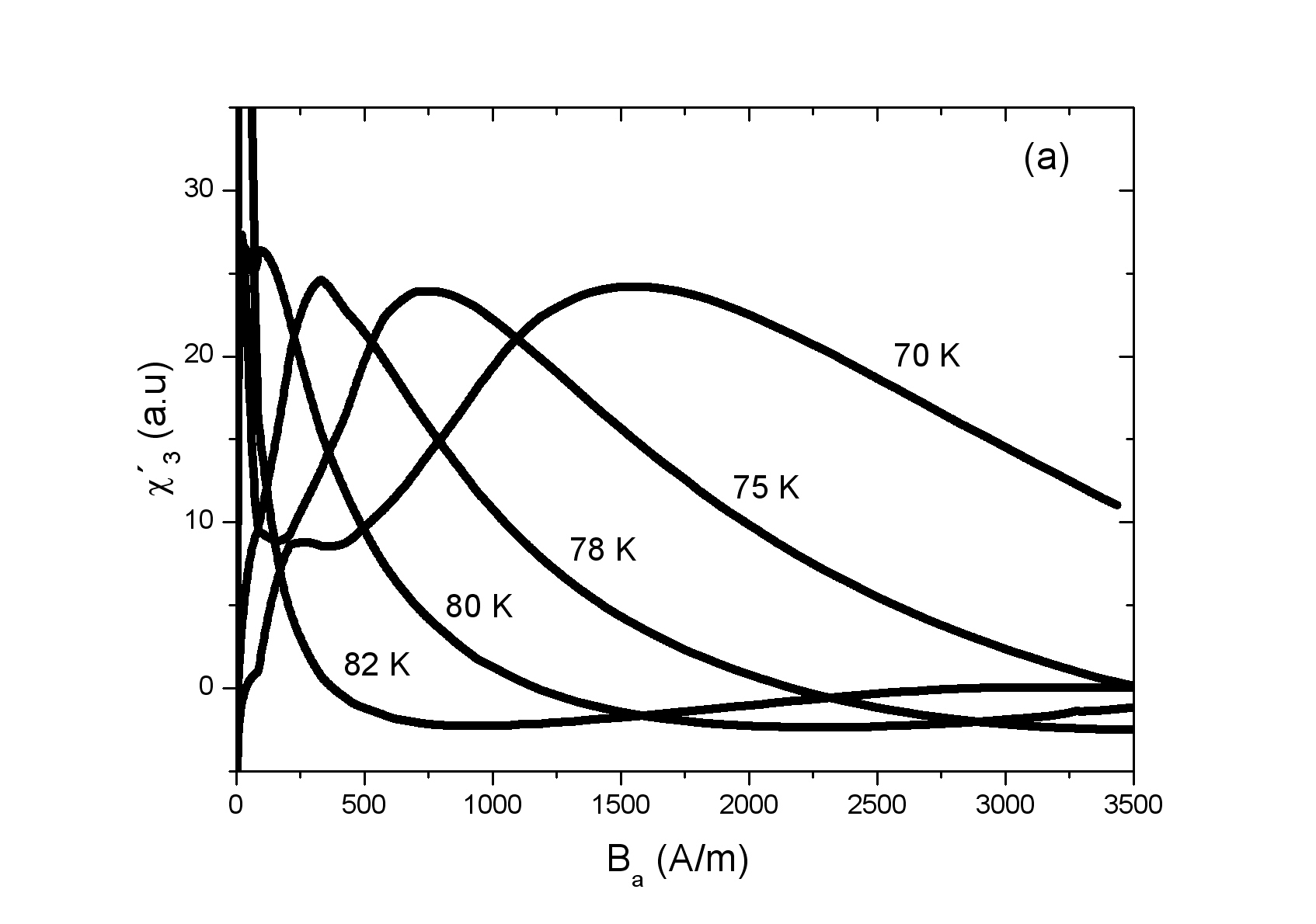} \includegraphics[width=6.5cm]{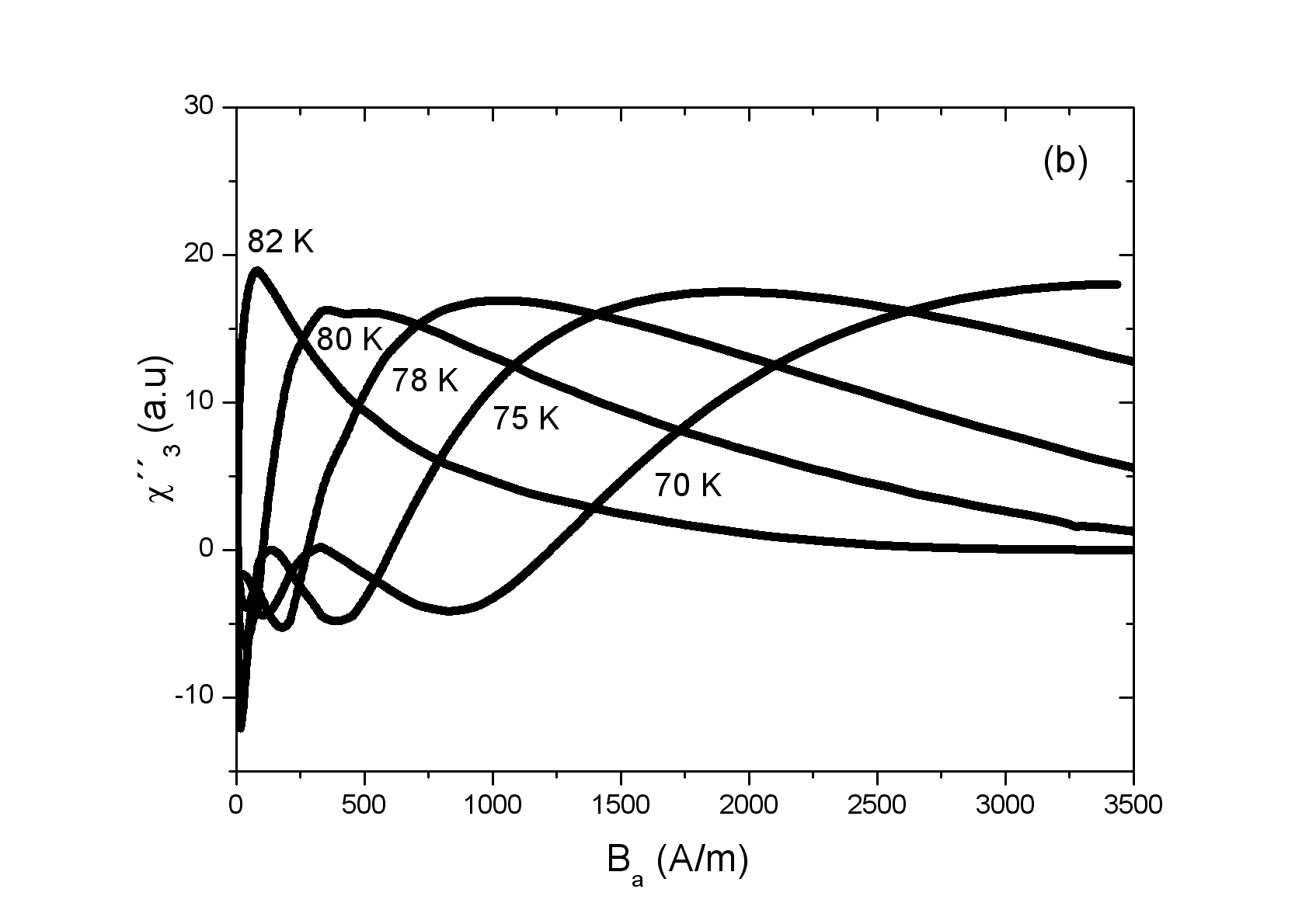} \includegraphics[width=6.5cm]{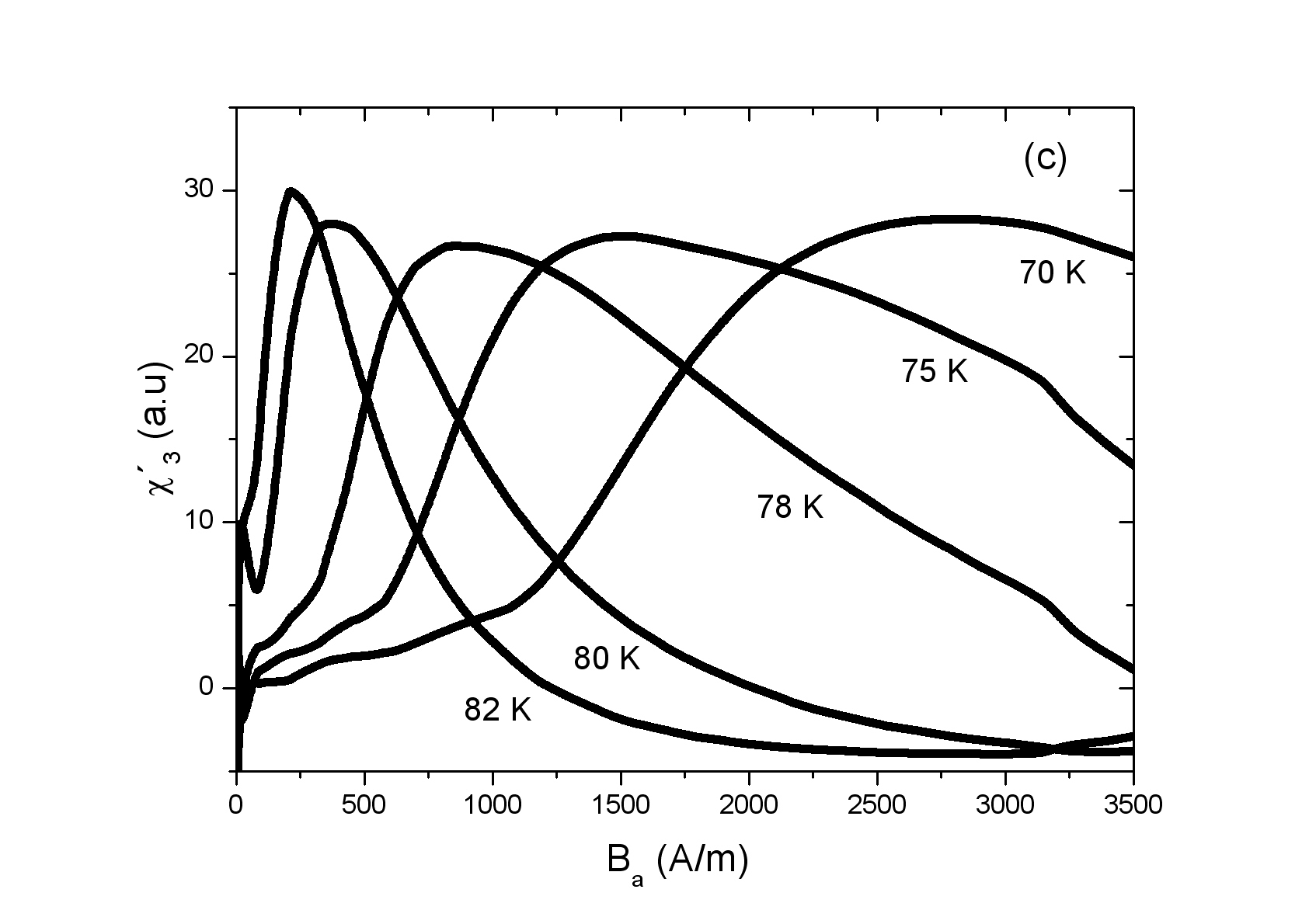} \includegraphics[width=6.5cm] {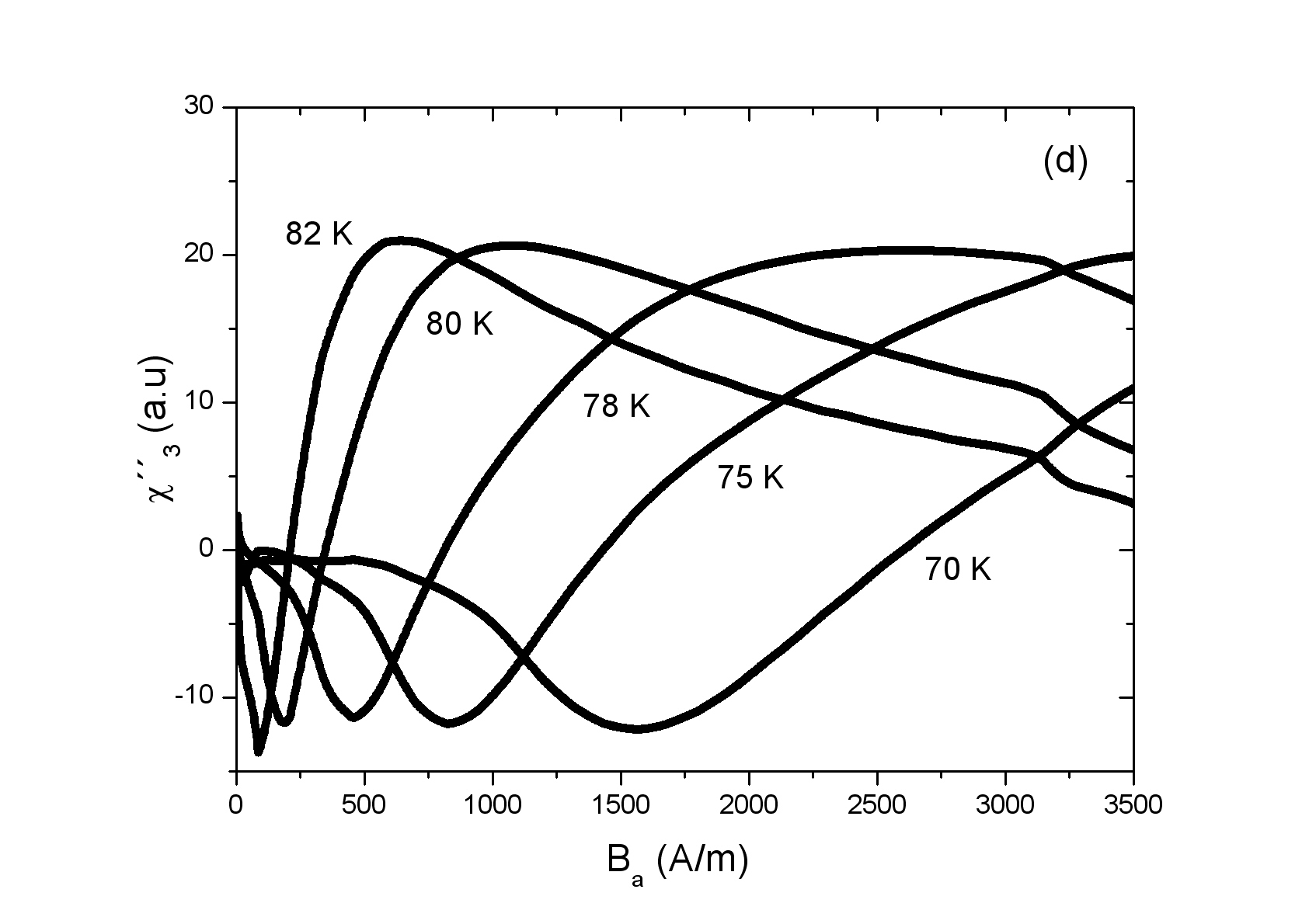} \includegraphics[width=6.5cm]{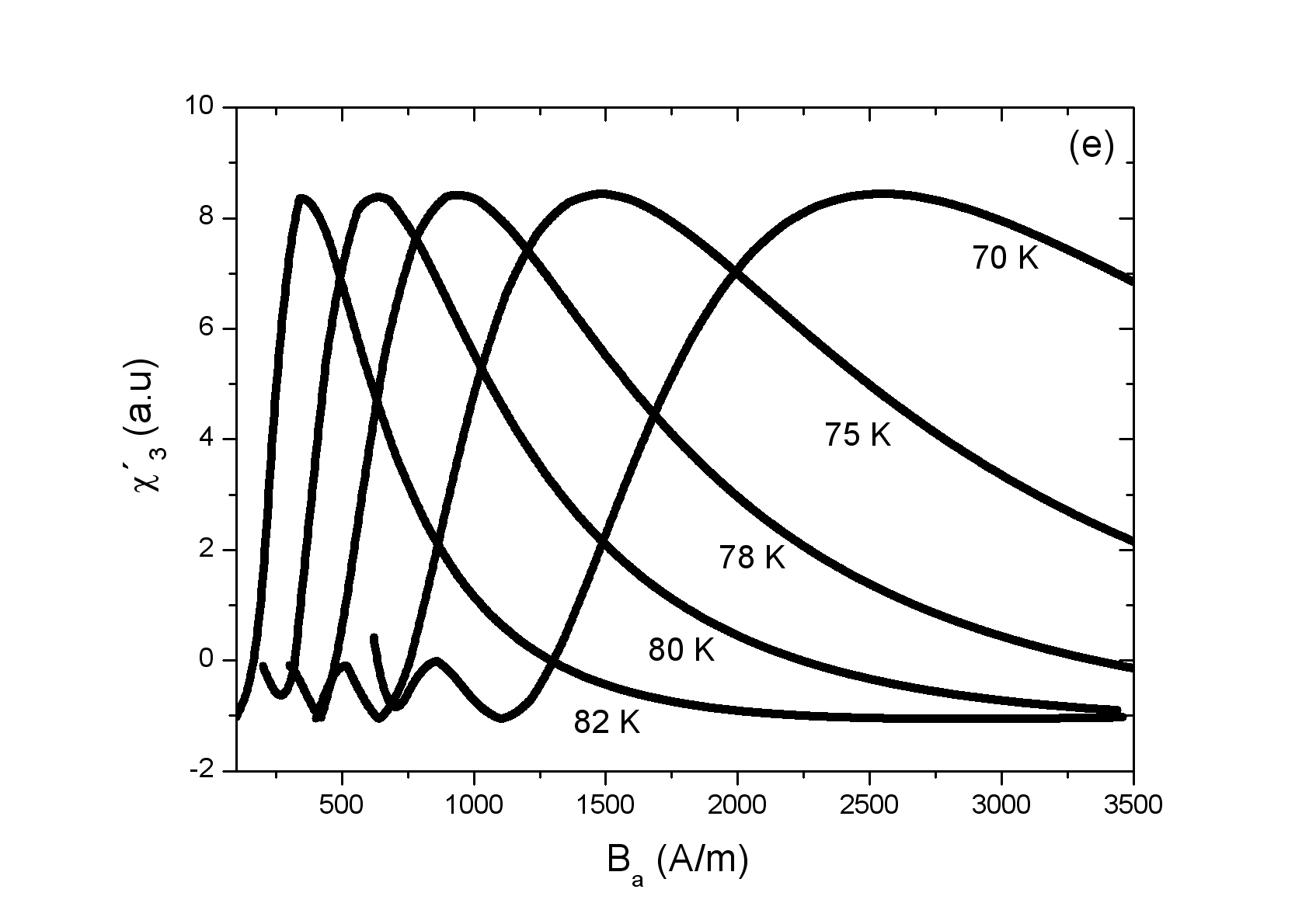} \includegraphics[width=6.5cm]{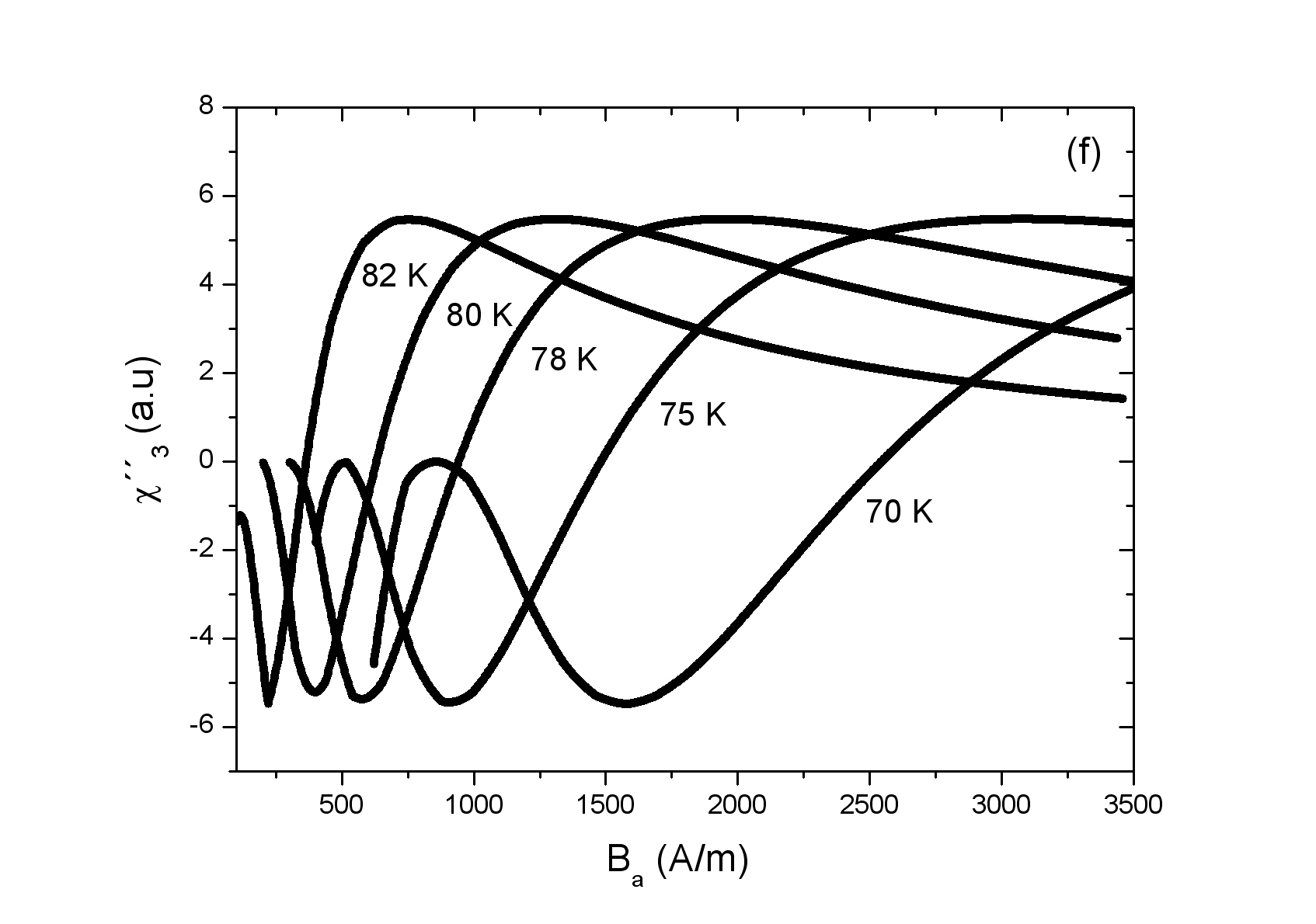}
\includegraphics[width=6.5cm]{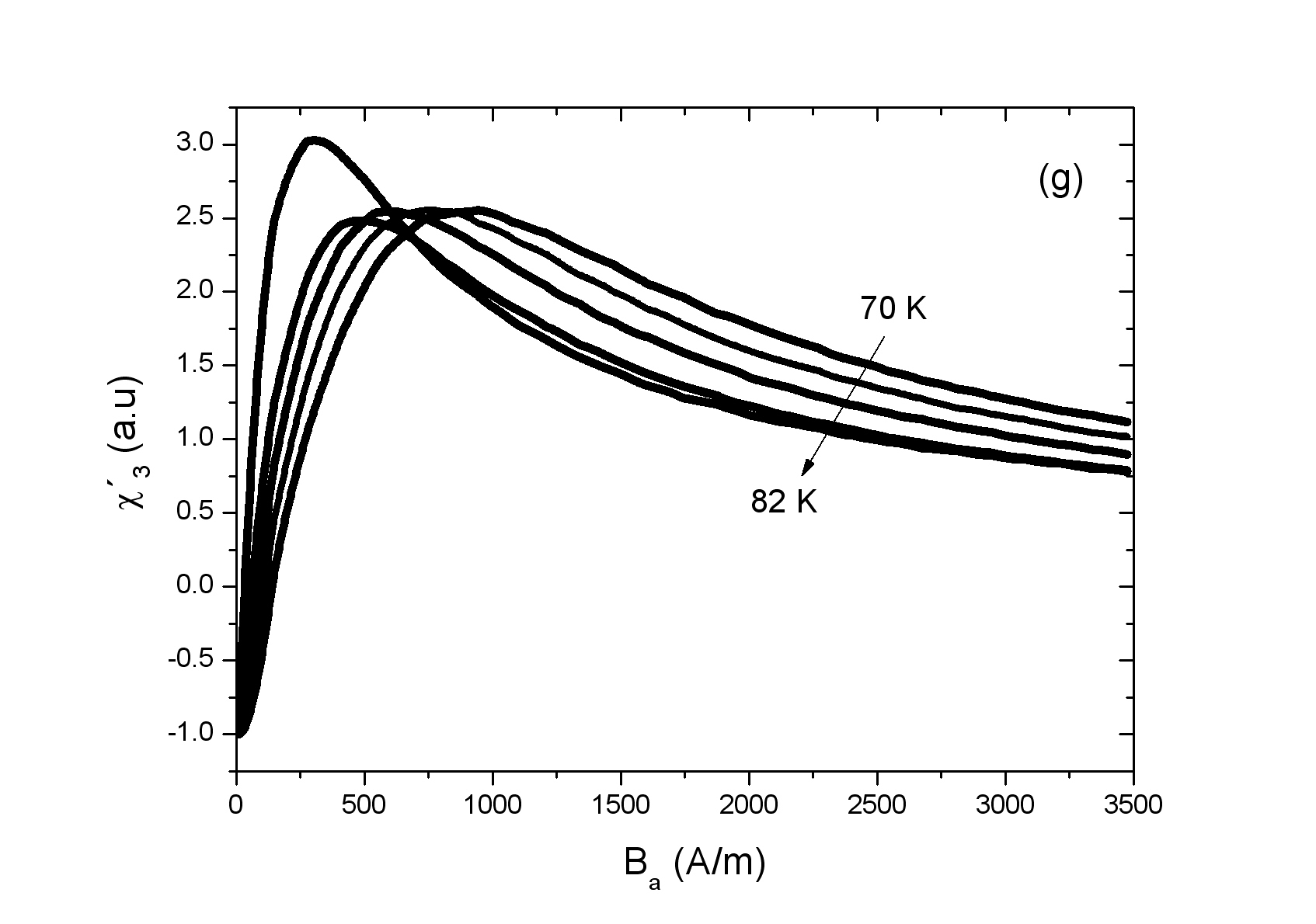} \includegraphics[width=6.5cm]{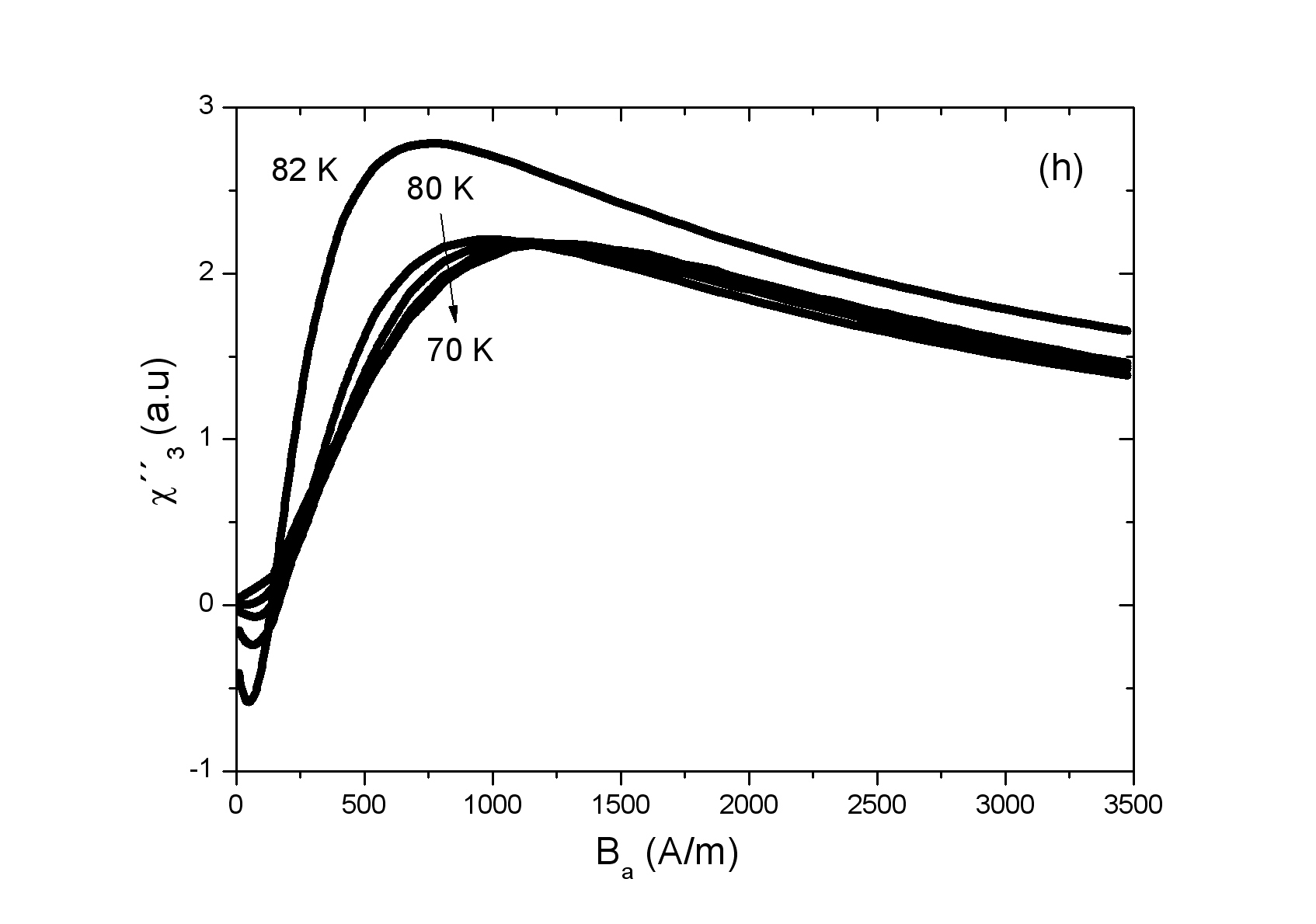}
\caption{Third harmonic of ac susceptibility response as function of applied field for the whole set of temperatures. (a) and (b): experimental curves for sample M22. (c) and (d): experimental curves for sample M23. (e) and (f): Ishida-Mazaki calculations for M23. (g) and (h): Brandt numerical calculations for M23. Left panels for real components. Right panels for imaginary components. As temperature lowers the main peak shifts to higher fields. The Brandt model qualitatively reproduces the shape of the experimental curves but the peak does not shift proportionally to the temperature changes.}
\label{third}
\end{center}
\end{figure*}

The field dependence for the third harmonic of susceptibility is shown Fig. \ref{third}. It is to be seen that the peaks of sample M22 are shifted towards the left for a given temperature in comparison with M23 [see Fig. \ref{third} a-d]. The physics involved in these curves is as follows. If the only mechanism governing the flux dynamics is pinning then for sufficiently low temperatures the vortices are strongly pinned and the maximum current density is achieved. As temperature increases vortices star to depin and the current density is significantly reduced. For temperatures close to $T_c$ the sample goes to the critical state unless the superconducting state is lost. Thus the slight difference in the position of the peaks for the two samples is due mainly to the lower $T_c$ of M22. This means that for a given field and temperature sample M22 goes faster into the mixed stated than sample M23.

Notice also that as the temperature decreases the transition in $\chi'_3$ and $\chi''_3$ broads. Moreover, the peak height in both components slightly decreases for temperatures far from $T_c$ but there is a great augmentation close to $T_c$. The increase in the magnitudes of the peaks reflects the increase in the nonlinearity in the magnetization, including hysteresis, as a function of the applied field. Another contribution to the increase of the peak comes from the field dependence of the magnetic susceptibility [see equation \eqref{sunsus}]. As the amplitude of the applied field tends to zero the components of the magnetic susceptibility diverge.

\begin{figure*}[htp]
\begin{center}
\includegraphics[width=6.5cm]{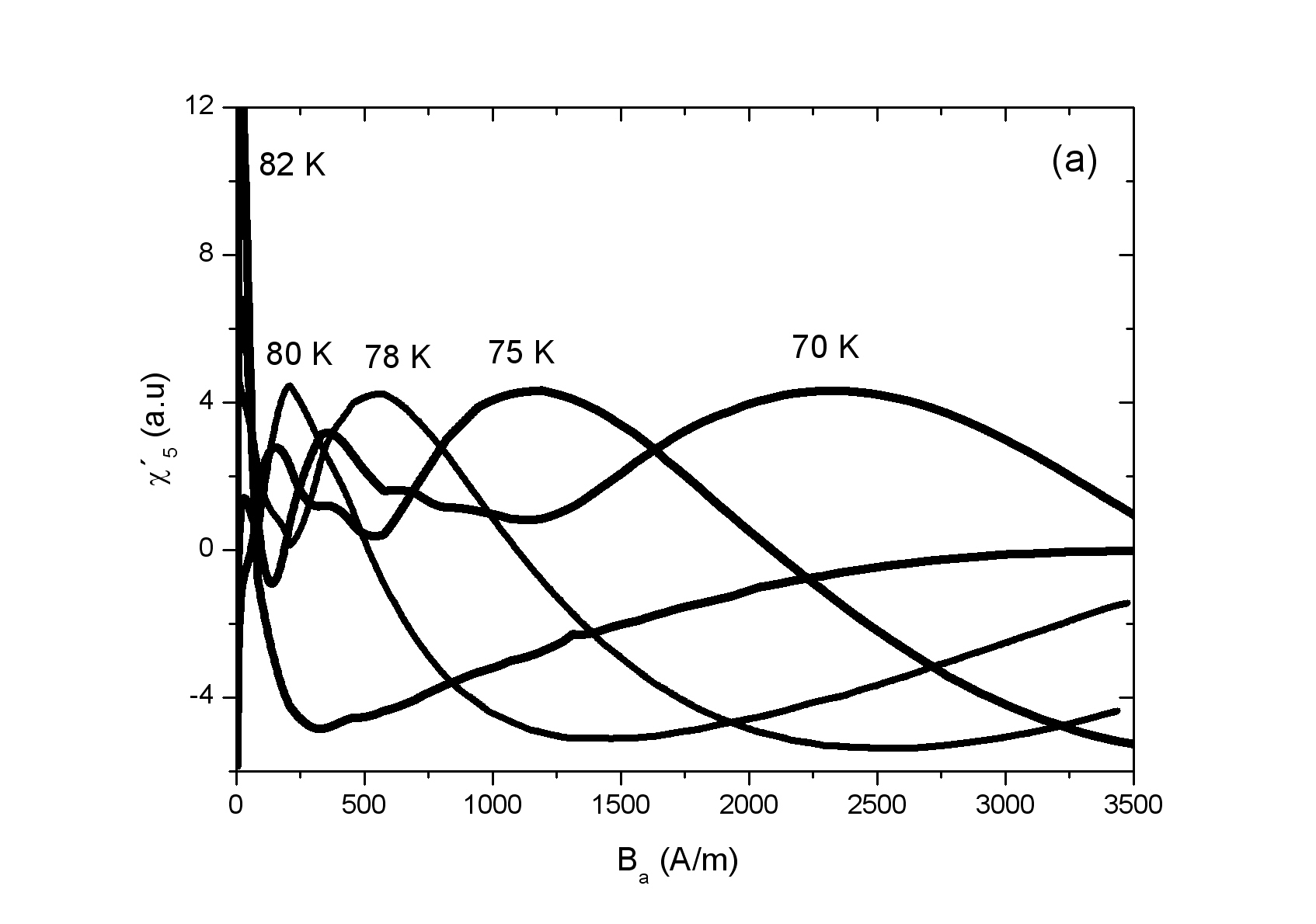} \includegraphics[width=6.5cm]{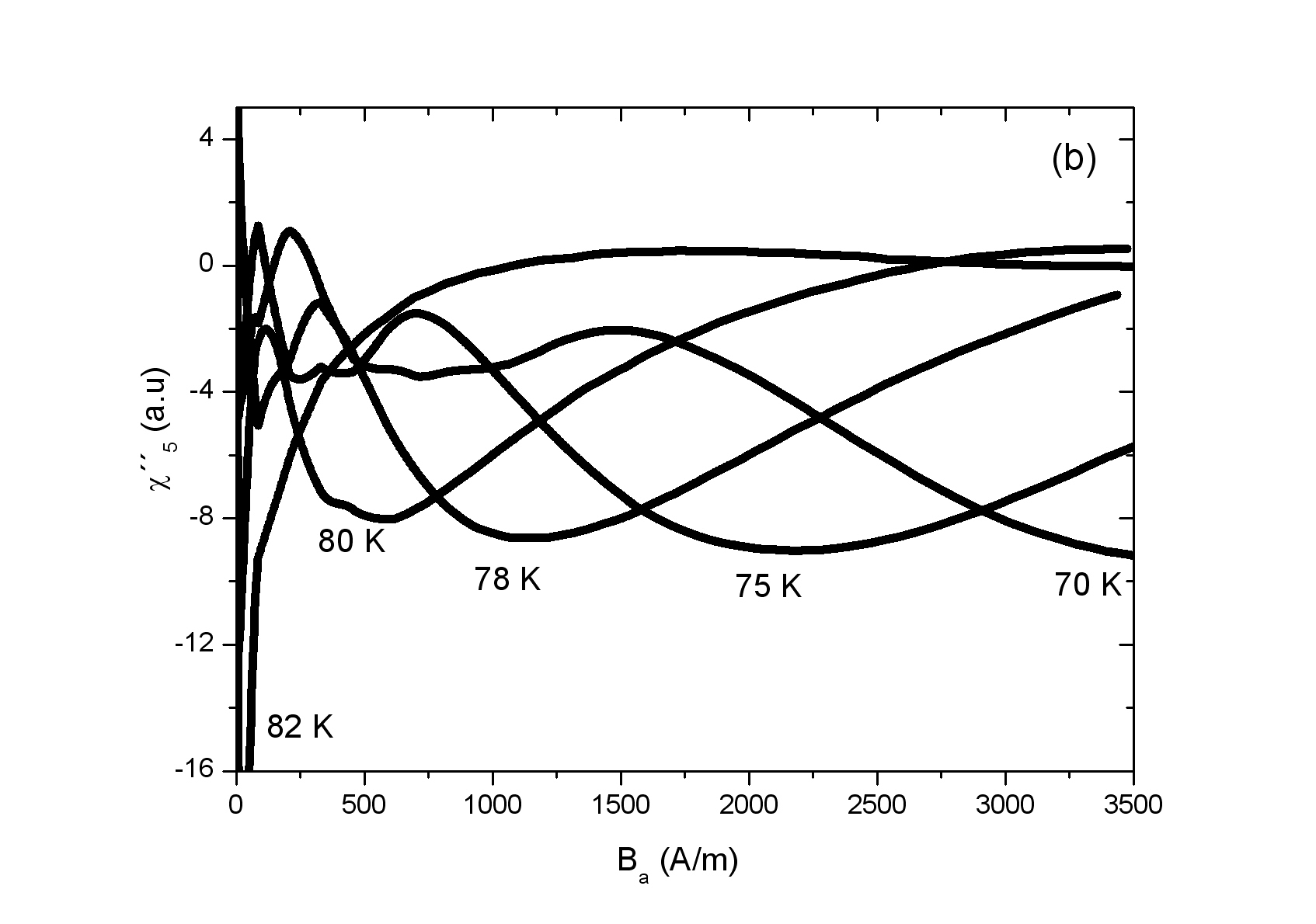} \includegraphics[width=6.5cm]{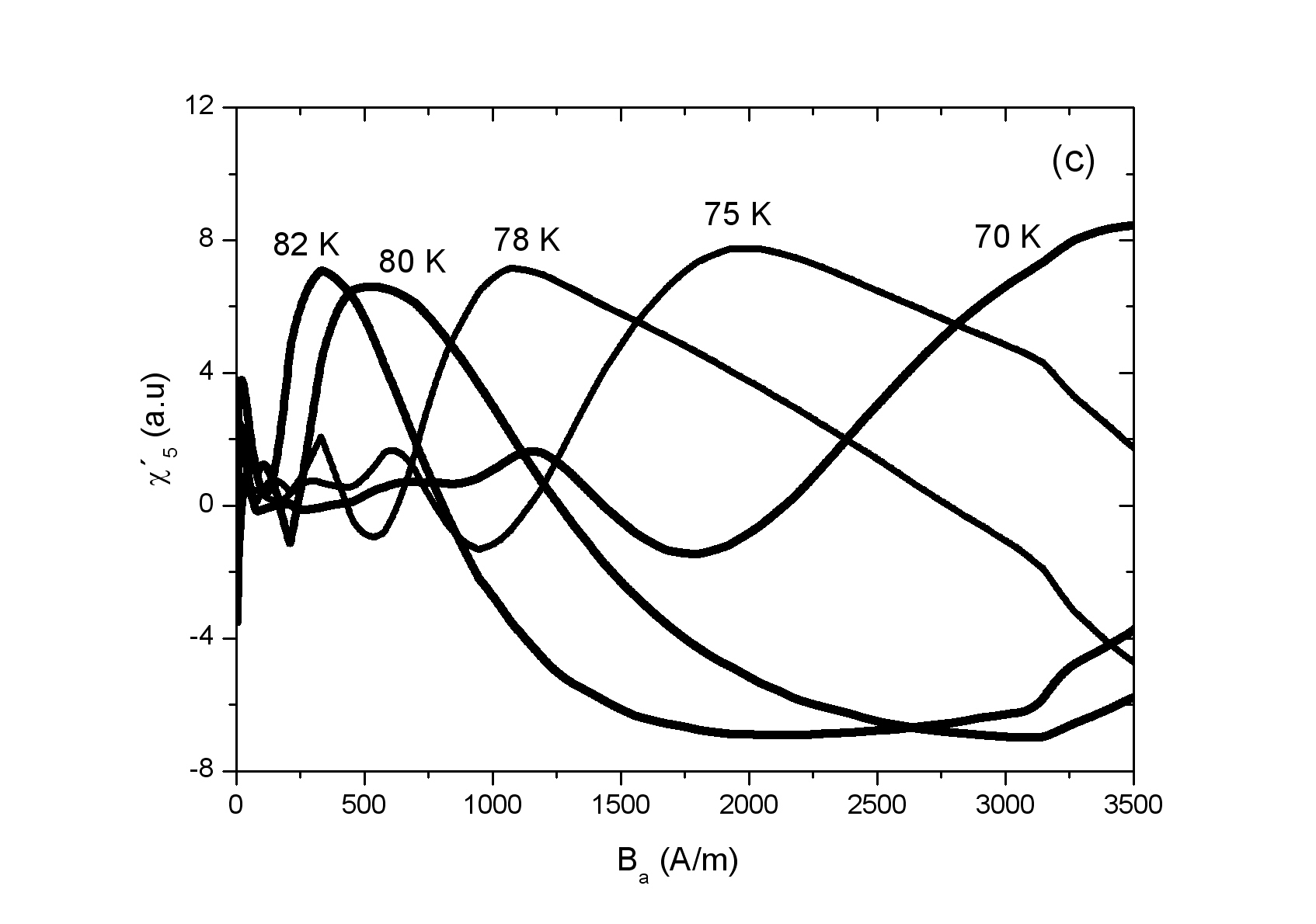} \includegraphics[width=6.5cm]{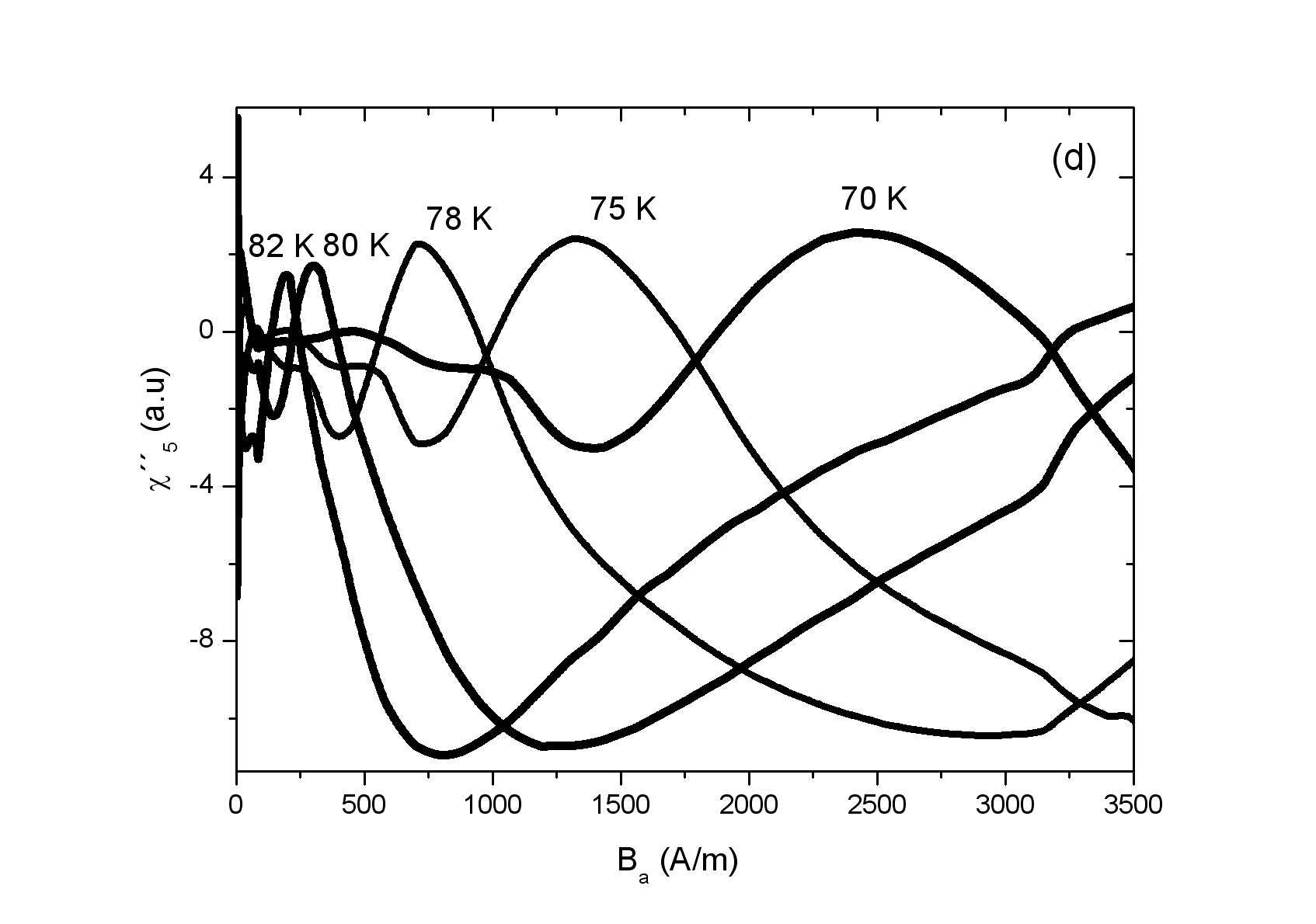} \includegraphics[width=6.5cm]{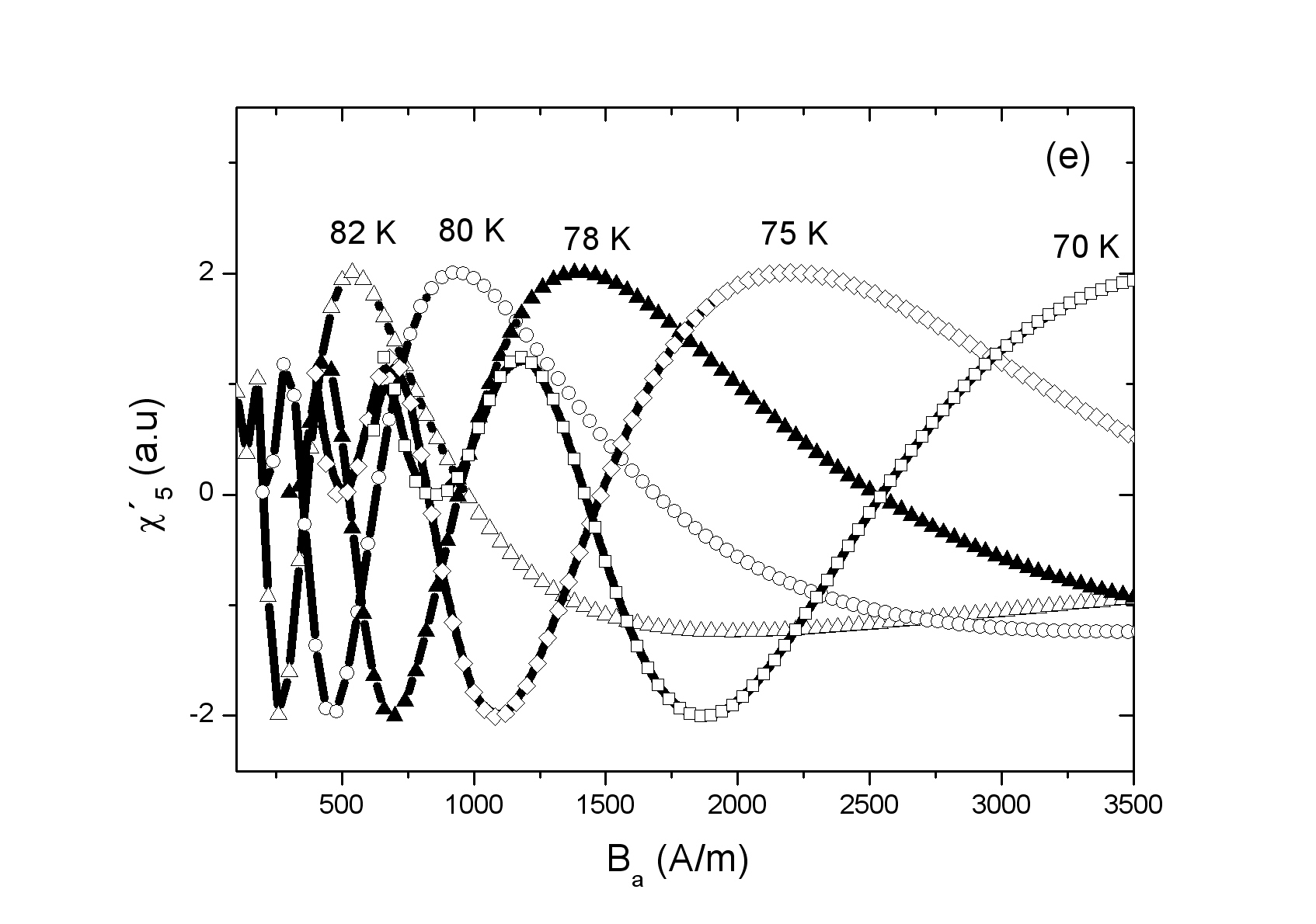} \includegraphics[width=6.5cm]{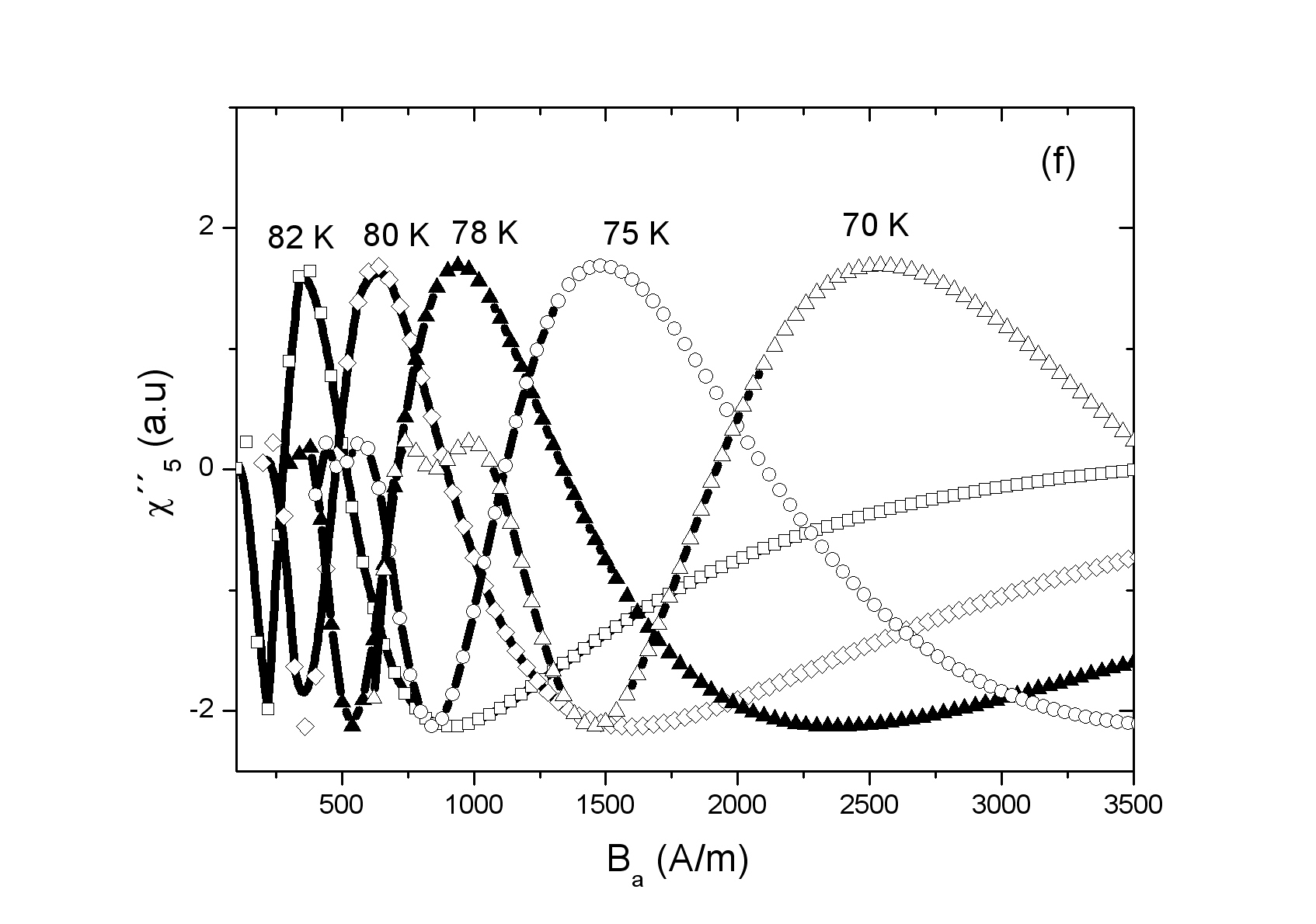}
\includegraphics[width=6.5cm]{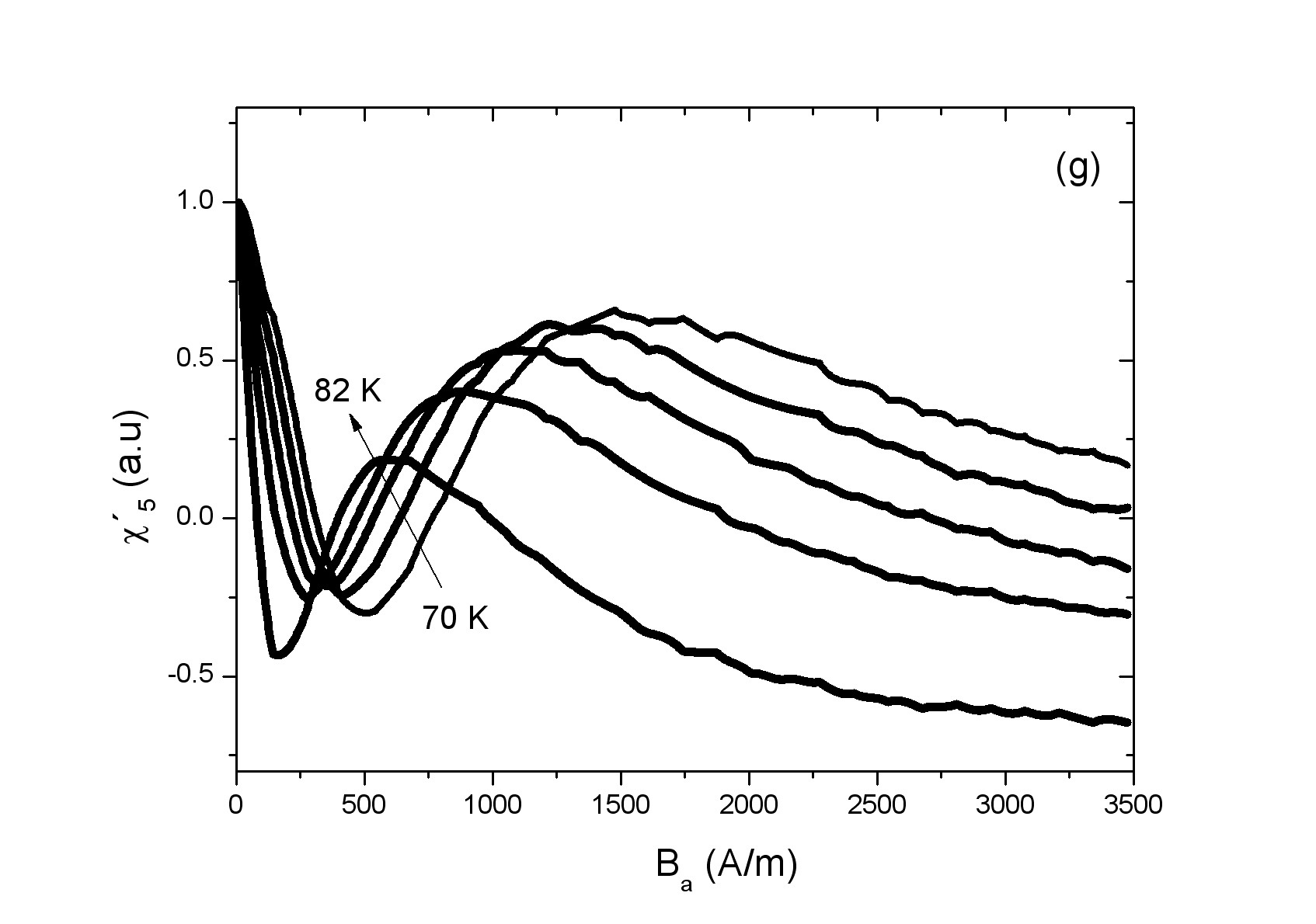} \includegraphics[width=6.5cm]{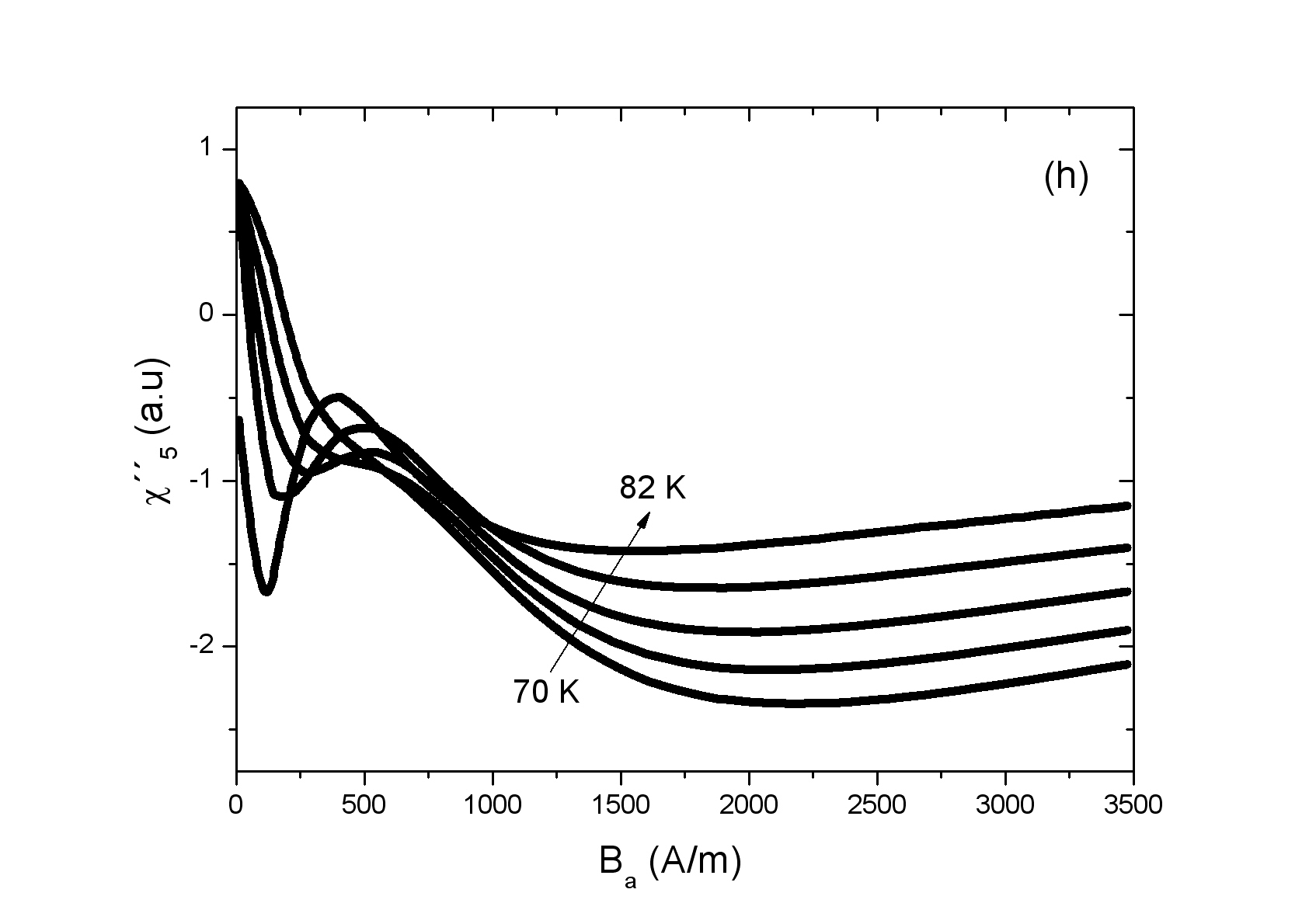}
\caption{Fifth harmonic of ac susceptibility response as function of applied field for the whole set of temperatures. (a) and (b): experimental curves for sample M22. (c) and (d): experimental curves for sample M23. (e) and (f) Ishida-Mazaki calculations for M23. And (g) and (h) Brandt calculations for M23. Left panels for real components. Right panels for imaginary components. As temperature lowers the main peak shifts to higher fields. The Brandt model shows a higher deviation in the agreement with the experimental curves. The peak also does not shift proportionally to the temperature changes.}
\label{fifth}
\end{center}
\end{figure*}
For comparison with the experimental data we have plotted the theoretical results for the third harmonic response for M23 in Figure \ref{third} e-h. The curves for M22 (not shown) only differ by a scaling factor that causes a reduction in the broadening but the curve structure is of same nature as that of M23. It is evident that the analytical calculations are in better agreement with the experimental results. Based on these results the best fitting values of  $J_{c0}$ for the Ishida-Mazaki model were $2 \times 10^{11}\, \textrm{A}/\textrm{m}^2$ and $8 \times 10^{11}\, \textrm{A}/\textrm{m}^2$ for M22 and M23, respectively. Furthermore, as can be seen from this figure the amplitude of the curves remains constant; this is attributed to the fact that the magnetic susceptibility \eqref{susishin} does not contemplate the variation of the amplitude of the applied magnetic field $B_a$. 

On the contrary, the numerical calculations reproduce better the amplitude of the experimental curves but the peak does not shift proportionally to the temperature changes  [Fig. \ref{third} g-h]. Also the curves do not fall off as fast as the experimental ones. These effects are probably due to the simplification realized in the power law relation \eqref{ej2} which cannot account for the temperature changes in the same proportion. This may be so, since the model we used only evaluates the resistivity as function of the current density which in turn is function of the vortex mobility in the film \cite{brandt11,brandt13,brandt14}. This may indicate that the model is not evaluating some parameter which is important to reproduce the temperature shift.

In Figure \ref{fifth} we plot the fifth component of susceptibility for the real and imaginary components for the two samples as well as the analytical and numerical calculations as function of the temperature and the applied field $B_{a}$. For the same reasons as in the third harmonic we have only presented the theoretical curves for M23. 

Here again we can observe that for temperatures close to $T_c$ the peak height greatly increases for the sample M22. Also as the temperature goes down from $T_c$ the curves become broad and the peak height is more stable. For comparison, the broadening for the curves of M23 is appreciably wider than that of M22 and the peak height, as the temperature goes down, increases. The plots also exhibit an oscillatory behavior for low magnetic fields. The analysis of the model has shown that this is mainly caused by the abrupt variations of the magnetic susceptibility as the applied field $B_a$ goes to zero through the variable $\alpha$. From equation \eqref{eqishi} we can clearly evidence that  for $B_a\ll B_m$ the function $\alpha$ varies very rapidly. Thus we can confirm that the overall structure of the data is in excellent agreement with the analytical model. 

The Brandt model, on the other hand, can only reproduce the overall structure of the experimental curves [see Fig. \ref{fifth} g-h]. Whilst, at low magnetic fields, the oscillatory behavior is not present. Notice that the curves do not shift proportionally to the temperature changes. Also the peak height of the real component increases as the temperature increases. In order to verify the cause of these discrepancies we tried several temperature dependences of $J_c$. Similarly, we manipulate the value of the creep parameter up to $m=100$, this manipulation is equivalent to a variation of the activation energy $U_c$. By following this procedure we are, in fact, inducing the model into the flux creep regime and the critical state, nevertheless, no new features appeared. These tests also support our previous arguments given for the third harmonic that the unexplainable effects may imply that the power law \eqref{ej2} for the current-voltage characteristics in the flux creep is not valid for the range of temperatures treated in this experiment. We have seen that the whole structure of even harmonics is well reproduced but perhaps a change in the material law \cite{aruna}, or a magnetic field dependence of $J_c$ or $U_c$ may take place \cite{moreno,wen}. However, in this article we did not include this models since there are parameters that we cannot measure, like, for instance, the characteristic length $\xi$ of the flux lattice line. 

\section{Conclusions}
\label{con}

Higher-order harmonics of the complex ac magnetic susceptibility have been measured for two superconducting thin films. We have compared the experimental results with theoretical ones which were based on the Ishida-Mazaki and the Brandt models. We have shown that the overall behavior of the analytical model is in excellent agreement with the experimental data. Based on this vision we may conclude that our thin films behave more as network of superconducting layers coupled by Josephson weak links.

On the other hand, although the overall structure of the magnetic behavior was reproduced by the Brandt model, it could not explain the peak shift as function of temperature. We attribute this failure to the simplification that we have made of the material law \eqref{matlaw}. Alternative approaches such as magnetic field dependence of the critical current density and activation energy must be considered to account for the lack of agreement with the experimental results.

\section*{Acknowledgements}
The authors acknowledge a CONACYT grant.

\end{document}